\documentclass[12pt]{article}
\pdfoutput=1 % for pdflatex
\usepackage{amsmath,amssymb,amsthm}
\usepackage{txfonts,bm,pifont}
\usepackage{graphicx}
\usepackage[pdfencoding=auto,bookmarks=true,bookmarksnumbered=true,colorlinks=true]{hyperref}
\usepackage{float}
\theoremstyle{plain}
\newtheorem{thm}{Theorem}[section]

\newtheorem{lemma}[thm]{Lemma}

\newtheorem{rem}[thm]{Remark}
\makeatletter
\newdimen\LENB \newdimen\LENW \newdimen\THI 
\newdimen\LENWH \newdimen\LENTOT \newcount\N 
\def\vbrknlnele#1#2#3{
  \LENB=#1pt \LENW=#2pt \THI=#3pt
  \LENWH=\LENW \divide\LENWH by 2
  \LENTOT=\LENB \advance\LENTOT by \LENW
  \vbox to \LENTOT{
    \vbox to \LENWH{}
    \nointerlineskip
    \vbox to \LENB{\hbox to \THI{\vrule width \THI height \LENB}}
    \nointerlineskip
    \vbox to \LENWH{}
  }}

\def\vbrknln#1{
  \N=#1
  \vcenter{
    \vbox{
      \loop\ifnum\N>0
        \vbox to 4pt{\vbrknlnele{2}{2}{0.1}}
        \nointerlineskip
        \advance\N by -1
      \repeat
  }}}

\def\vbl#1{\hskip-5pt \vbrknln{#1} \hskip-5pt}

\def\hbrknlnele#1#2#3{
  \LENB=#1pt \LENW=#2pt \THI=#3pt
  \LENTOT=\LENB \advance\LENTOT by \LENW
  \vcenter{
    \vbox to \THI{
      \hbox to \LENTOT{
        \hfil
        \vrule width \LENB height \THI
        \hfil}
  }}}

\def\hblele{\hbrknlnele{2}{2.2}{0.1}}

\def\hblfil{\cleaders\hbox{$ \m@th \mkern1mu \hblele \mkern1mu
$}\hfill} 
\makeatother
\begin{document}
\begin{center}
\begin{Large}
dNLS Flow on Discrete Space Curves\\[5mm]
\end{Large}
\begin{normalsize}
Sampei {\sc Hirose}\\
Center for Promotion of Educational Innovation,
Shibaura Institute of Technology\\
307 Fukasaku, Minuma-ku, Saitama 337-8570, Japan\\
e-mail: hirose3@shibaura-it.ac.jp\\[2mm]
Jun-ichi {\sc Inoguchi}\\
Institute of Mathematics, University of Tsukuba\\
Tsukuba 305-8571, Japan\\
e-mail: inoguchi@math.tsukuba.ac.jp\\[2mm]
Kenji {\sc Kajiwara}\\
Institute of Mathematics for Industry, Kyushu University\\
744 Motooka, Fukuoka 819-0395, Japan\\
e-mail: kaji@imi.kyushu-u.ac.jp\\[2mm]
Nozomu {\sc Matsuura}\\
Department of Applied Mathematics, Fukuoka University\\
Nanakuma 8-19-1, Fukuoka 814-0180, Japan\\
e-mail: nozomu@fukuoka-u.ac.jp\\[2mm]
Yasuhiro {\sc Ohta}\\
Department of Mathematics, Kobe University\\
Rokko, Kobe 657-8501, Japan\\
e-mail: ohta@math.sci.kobe-u.ac.jp
\end{normalsize}
\end{center}
\begin{abstract}
The local induction equation, or the binormal flow on space curves is a well-known model of
deformation of space curves as it describes the dynamics of vortex filaments, and the complex
curvature is governed by the nonlinear Schr\"odinger equation (NLS). In this paper, we present its
discrete analogue, namely, a model of deformation of discrete space curves by the discrete nonlinear
Schr\"odinger equation (dNLS). We also present explicit formulas for both NLS and dNLS flows in
terms of the $\tau$ function of the 2-component KP hierarchy.
\end{abstract}
\section{Introduction}
The local induction equation (LIE)
\begin{equation}\label{eqn:local_induction}
\frac{\partial\gamma}{\partial t} = \frac{\partial\gamma}{\partial x}\times \frac{\partial^2\gamma}{\partial x^2},
\end{equation}
is one of the most important models of deformation of space curves, where
$\gamma(x,t)\in\mathbb{R}^3$ is a smooth space curve parametrized by the arc-length $x$ and $t$ is a
deformation parameter \cite{Hasimoto:JFM1972,Lamb:JMP1977,Nakayama_Segur_Wadati:PRL}.  In a physical
setting, it describes the dynamics of vortex filaments driven by the self-induction in the inviscid
fluid under the local induction approximation \cite{Hasimoto:JFM1972}.

It is well-known that if $\gamma$ obeys LIE, then the curvature and the torsion, or equivalently,
the complex curvature of $\gamma$ solves the {\em nonlinear Schr\"odinger equation} (NLS) which is
one of the most typical equations in the integrable systems.  To show this, we use the Frenet frame
$\Phi=\Phi(x,t)=[T(x,t), N(x,t),B(x,t)]\in\mathrm{SO}(3)$, where $T$, $N$, $B$ are the tangent, the
normal, and the binormal vectors defined by
\begin{equation}
 T=\gamma',\quad 
N=\frac{\gamma''}{\left|\gamma''\right|},\quad
B=T\times N,\quad ' = \frac{\partial}{\partial x},
\end{equation}
respectively. Note that it follows that $ |T|=\left|\gamma'\right|=1$ since $x$ is the
arc-length. Then we have the {\em Frenet-Serret formula}
\begin{equation}\label{eqn:Frenet-Serret}
\frac{\partial \Phi}{\partial x} 
= \Phi L,\quad L=\left[\begin{array}{ccc} 0& -\kappa& 0\\ \kappa & 0 & -\lambda \\ 0 &\lambda &0\end{array}\right],
\end{equation}
where $\kappa = \left|\gamma''\right|$ and $\lambda = -\langle B',N\rangle$ are the curvature and
the torsion, respectively.  In this setting, LIE \eqref{eqn:local_induction} is expressed as the
deformation by the binormal flow
\begin{equation}\label{eqn:continuous_binormal_flow}
 \frac{\partial\gamma}{\partial t} = \kappa B,
\end{equation}
and the corresponding deformation equation of the Frenet frame is given by
\begin{equation}\label{eqn:continuous_deformation_Frenet_frame}
 \frac{\partial\Phi}{\partial t} 
= \Phi M,\quad M=
\begin{bmatrix}
0 & \kappa \lambda & - \kappa'\\
- \kappa \lambda & 0 & - \frac{\kappa''}{\kappa} + \lambda^2\\
\kappa' & \frac{\kappa''}{\kappa} - \lambda^2 & 0
\end{bmatrix}.
\end{equation}
The compatibility condition of the system of linear partial differential equations
\eqref{eqn:Frenet-Serret} and \eqref{eqn:continuous_deformation_Frenet_frame}
\begin{equation}
 \frac{\partial L}{\partial t} -  \frac{\partial M}{\partial x} = LM - ML,
\end{equation}
yields
\begin{equation}\label{eqn:NLS_kappa}
\frac{\partial\kappa}{\partial t}
= - 2 \frac{\partial \kappa}{\partial x} \lambda - \kappa \frac{\partial\lambda}{\partial x},\quad
\frac{\partial \lambda}{\partial t}
= \frac{\partial }{\partial x}\left(\frac{\kappa''}{\kappa}+ \frac{\kappa^2}{2} - \lambda^2\right).
\end{equation}
Introducing the {\em complex curvature} $u=u(x,t)\in\mathbb{C}$ by the {\em Hasimoto transformation} \cite{Hasimoto:JFM1972}
\begin{equation}\label{eqn:Hasimoto_transform}
 u = \kappa e^{\sqrt{-1}\Lambda},\quad \Lambda = \int^x \lambda\,dx,
\end{equation}
we see that $u$ satisfies NLS
\begin{equation}\label{eqn:NLS}
\sqrt{-1}\frac{\partial u}{\partial t} + \frac{\partial^2u}{\partial x^2} + \frac{1}{2}|u|^2 u = 0.
\end{equation}
Also, one can show that this deformation is isoperimetric, namely $|\gamma'|=1$ for all $t$.

Discretization of curves and their deformations preserving underlying integrable structures is an
important problem in the discrete differential geometry.  For example, the isoperimetric deformation
of plane discrete curves described by the discrete mKdV equation (dmKdV) has been studied in
\cite{IKMO:KJM,IKMO:MEIS2013,Matsuura:IMRN}. For discrete space curves, the deformations by the
discrete sine-Gordon equation (dsG) and dmKdV has been studied in
\cite{Doliwa-Santini:dsG,IKMO:dmKdV_space_curve,IKMO:MEIS2013}, and the deformation by dNLS is
formulated in \cite{Hoffmann:dNLS,Pinkall:dNLS}.

The dsG and dmKdV describe torsion-preserving isoperimetric and equidistant deformation of the space
discrete curves with constant torsion. However, formulation of discrete deformation of space
discrete curves with varying torsion is a difficult problem. The only example so far is presented by
Hoffmann \cite{Hoffmann:dNLS_HM,Hoffmann:dNLS}, where he has claimed that composition of certain two
isoperimetric equidistant deformations can be regarded as a discrete analogue of LIE.  Also, it was
used for numerical simulation of fluid flow
\cite{Pinkall:dNLS,Weissmann-Pinkall:dNLS_simulation}. This formulation uses quarternions and its
geometric meaning is clear, but description of the deformation parameters in terms of the complex
curvature, thus the relation to dNLS are rather indirect.

In this paper, we present a formulation of the dNLS flow on discrete space curves from different
approach; the deformation of curves is expressed in terms of the discrete Frenet frame with the
coefficients given by the curvature and torsion of the curves explicitly. In this approach, dNLS
arises as the equation governing the complex curvature of curves, which is the same as the case of
smooth curves. Based on this formulation, we present explicit formulas for the NLS flow for smooth
curves and the dNLS flow to discrete curves in terms of $\tau$ functions of the two-component KP
hierarchy by applying the theory of integrable systems. We expect that our dNLS flow can be an
alternative to Hoffmann's formulation when it is used to simulate the dynamics of fluids.
Also, explicit expression of the scheme and exact solutions may promote further development of
theoretical studies of discrete dynamics of discrete curves from both mathematical and physical point of view.
%%%%%%%%%%%%%%%%%%%%%%%%%%%%%%%%%%%%%%%%%%%%%%%%%%%%
%
%%%%%%%%%%%%%%%%%%%%%%%%%%%%%%%%%%%%%%%%%%%%%%%%%%%%
\section{dNLS flow on discrete space curves} 
Let $\gamma_n\in\mathbb{R}^3$ be a discrete space curve with
\begin{equation}
 |\gamma_{n+1}-\gamma_n| = \epsilon,
\end{equation}
where $\epsilon$ is a constant. We introduce the {\em discrete Frenet frame}
$\Phi_n=[T_n,N_n,B_n]\in\mathrm{SO}(3)$ by
\begin{equation}\label{eqn:discrete_Frenet_frame}
T_n = \frac{\gamma_{n+1}-\gamma_n}{\epsilon},\quad
B_n = \frac{T_{n-1}\times T_n}{\left|T_{n-1}\times T_n\right|},\quad
N_n = B_n\times T_n.
\end{equation}
Then it follows that the discrete Frenet frame satisfies the {\em discrete Frenet-Serret formula}
\begin{equation}\label{eqn:discrete_Frenet-Serret}
 \Phi_{n+1} = \Phi_n L_n,\quad L_n = R_1(-\nu_{n+1}) R_3(\kappa_{n+1}),
\end{equation}
where
\begin{equation}
 R_1(\theta) = 
\left[\begin{array}{ccc}
1 & 0          & 0          \\
0 & \cos\theta  & -\sin\theta\\
0 & \sin\theta  &  \cos\theta\\
\end{array}\right],\quad
 R_3(\theta) = 
\left[\begin{array}{ccc}
 \cos\theta  & -\sin\theta & 0\\
 \sin\theta  &  \cos\theta & 0\\
     0      &       0     & 1
\end{array}\right],
\end{equation}
and $\nu_n$, $\kappa_n$ are defined by
\begin{equation}
\begin{split}
& \langle T_{n-1},T_n\rangle =\cos\kappa_n,\quad  \langle B_{n},B_{n-1}\rangle =\cos\nu_n,\quad 
\langle B_{n},N_{n-1}\rangle =\sin\nu_n,\\
&\hskip60pt -\pi\leq \nu_n< \pi,\quad 0<\kappa_n<\pi. 
\end{split}
\end{equation}
In order to formulate a ``good'' discrete deformation (discretization of time), we resort to the
theory of discrete integrable systems to preserve integrable nature of the NLS flow
\eqref{eqn:continuous_binormal_flow}. As a discrete analogue of NLS \eqref{eqn:NLS}, we consider
\begin{equation}\label{eqn:dNLS}
\begin{split}
&   {\textstyle \left({\scriptstyle  \sqrt{-1}}\frac{\epsilon^2}{\delta} - 1\right)}u_{n}^{m+1} 
-  {\textstyle \left({\scriptstyle\sqrt{-1}}\frac{\epsilon^2}{\delta} + 1\right)}u_n^m
+ (u_{n+1}^m+u_{n-1}^{m+1})(1+\epsilon^2|u_n^m|^2)\Gamma_n^m=0 ,\\[2mm]
&\hskip60pt  \frac{\Gamma_{n+1}^m}{\Gamma_n^m} = \frac{1+\epsilon^2|u_n^m|^2}{1+\epsilon^2|u_n^{m+1}|^2},
\end{split}
\end{equation}
which we refer to as the {\em discrete nonlinear Schr\"odinger equation}
(dNLS) \cite{Ablowitz-Ladik:SiAM1976,Ablowitz-Ladik:SiAM1977,Hirota-Tsujimoto:App_book}.  Here,
$u_n^m\in\mathbb{C}$, $\Gamma_n^m\in\mathbb{R}$, $n$ is the space discrete variable which
corresponds to the label of vertices of discrete curves, $m$ is the discrete variable corresponding
to the step of deformation, $\epsilon$ and $\delta$ are constants which are the lattice intervals of
$n$ and $m$, respectively.  Moreover, $u_n^m$ is the complex discrete curvature defined by
\begin{equation}
 u_n^m = \frac{1}{\epsilon}\tan\frac{\kappa_n^m}{2} e^{\sqrt{-1}\Lambda_n^m},\quad \Lambda_{n}^m - \Lambda_{n-1}^m = -\nu_n^m,
\end{equation}
We impose the boundary condition as
\begin{equation}\label{eqn:dNLS_bc}
 u_n^m\to 0\ (n\to\pm\infty),\quad \Gamma_n^m\to \Gamma_{\pm\infty}\ (\text{const.})\ (n\to\pm\infty).
\end{equation}
Then one of the main statements of this paper is given as follows:
%%%%%%%%%%%%%%%%%%%%%%%%%%%%%%%%%%%%%%%%%%%%%%%%%%%%%%%%%%%%%%
%
%%%%%%%%%%%%%%%%%%%%%%%%%%%%%%%%%%%%%%%%%%%%%%%%%%%%%%%%%%%%%%
\begin{thm}[dNLS flow]\label{thm:discrete_deformation}\hfill\\
For a fixed $m$, let $\gamma_n^m\in\mathbb{R}^3$ be a discrete space curve satisfying
\begin{equation}
 |\gamma_{n+1}^m - \gamma_n^m | = \epsilon,
\end{equation}
and $\Phi_n^m=[T_n^m, N_n^m, B_n^m]\in\mathrm{SO}(3)$ be the discrete Frenet frame defined in
\eqref{eqn:discrete_Frenet_frame} satisfying the discrete Frenet-Serret formula
\begin{equation}\label{eqn:fulldiscrete_Frenet-Serret}
 \Phi_{n+1}^m = \Phi_n^m L_n^m,\quad L_n^m = R_1(-\nu_{n+1}^m)R_3(\kappa_{n+1}^m).
\end{equation}
%=\frac{1}{\epsilon}\tan\frac{\kappa_n^m}{2} e^{\sqrt{-1}\Lambda_n^m}
Let $u_n^m$ be a complex discrete curvature of $\gamma_n^m$. We determine $u_n^{m+1}$ by dNLS
\eqref{eqn:dNLS} under the boundary condition \eqref{eqn:dNLS_bc} and put
$u_n^{m+1}=\frac{1}{\epsilon}\tan\frac{\kappa_n^{m+1}}{2} e^{\sqrt{-1}\Lambda_n^{m+1}}$.  We define
a new curve $\gamma_n^{m+1}\in\mathbb{R}^3$ by
\begin{equation}\label{eqn:discrete_def_Frenet}
 \frac{\gamma_n^{m+1} - \gamma_n^m}{\delta} = \frac{2}{\epsilon^3}(P_n^m T_n^m + Q_n^m N_n^m + R_n^m B_n^m),
\end{equation}
\begin{equation}\label{eqn:discrete_coeffs_Frenet}
 \begin{split}
&P_n^m = \delta\left(-1 + \frac{\Gamma_n^m}{\cos^2\frac{\kappa_n^m}{2}}\right),\\
&Q_n^m = \delta\left[\tan\frac{\kappa_n^m}{2} - \tan\frac{\kappa_{n-1}^{m+1}}{2}
\cos(\Lambda_{n-1}^{m+1}-\Lambda_n^m)\frac{\Gamma_n^m}{\cos^2\frac{\kappa_n^m}{2}}\right],\\
& R_n^m= \epsilon^2\tan\frac{\kappa_n^m}{2} 
- \delta\tan\frac{\kappa_{n-1}^{m+1}}{2}\sin(\Lambda_{n-1}^{m+1}-\Lambda_n^m)\frac{\Gamma_n^m}{\cos^2\frac{\kappa_n^m}{2}}.
 \end{split}
\end{equation}
Suppose that $\Gamma_{\infty}$ and $\Gamma_{-\infty}$ are either $1$ or
$1+\frac{\epsilon^4}{\delta^2}$. Then, it follows that 
\begin{enumerate}
 \item $|\gamma_{n+1}^{m+1}- \gamma_n^{m+1}|=\epsilon$. Namely, $\gamma_n^{m+1}$ is an
isoperimetric deformation of $\gamma_n^m$.
 \item $u_n^{m+1}$ gives the complex discrete curvature of $\gamma_n^{m+1}$.
\end{enumerate}
\end{thm}
\begin{figure}[ht]
 \begin{center}
\includegraphics[bb=0 0 150 380,scale=0.25]{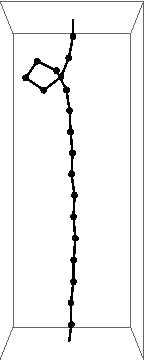}\hskip40pt
\includegraphics[bb=0 0 150 380,scale=0.25]{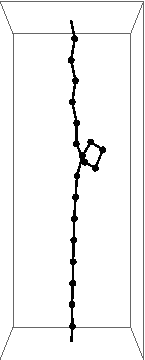}\hskip40pt
\includegraphics[bb=0 0 150 380,scale=0.25]{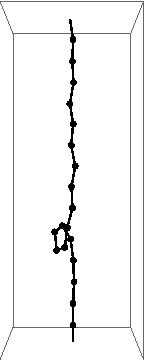}\hskip40pt
\includegraphics[bb=0 0 150 380,scale=0.25]{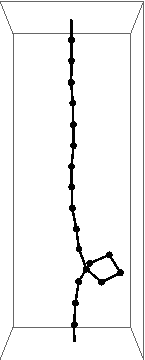}
\end{center}
\caption{Numerical simulation of dNLS flow. }
\end{figure}
%%%%%%%%%%%%%%%%%%%%%%%%%%%%%%%%%%%%%%%%%%%%%%%%%%%%%%%%%%%%%%
%
%%%%%%%%%%%%%%%%%%%%%%%%%%%%%%%%%%%%%%%%%%%%%%%%%%%%%%%%%%%%%%
\begin{rem}\hfill
\begin{enumerate}
 \item The deformation \eqref{eqn:discrete_coeffs_Frenet} is {\em not} an equidistant deformation in
contrast with the deformation described by dmKdV \cite{IKMO:dmKdV_space_curve}. In fact,
one can show that
\begin{equation}\label{eqn:discrete_displacement_distance}
 \left|\frac{\gamma_n^{m+1}-\gamma_n^m}{\delta}\right|^2 
= \frac{4}{\epsilon^2}\left(-1 + \frac{\Gamma_n^{m}}{\cos^2\frac{\kappa_n^m}{2}}\right).
\end{equation}
Equation \eqref{eqn:discrete_displacement_distance} also implies that the solution of dNLS \eqref{eqn:dNLS}
should satisfy the condition $\Gamma_n^{m}\geq \cos^2\frac{\kappa_n^m}{2}$ in order
to be consistent with the curve deformation. 
Note that this property does not contradict with
Hoffmann's formulation where the deformation is constructed as composition of two isoperimetric and
equidistant deformations.
 \item Continuous limit with respect to time can be simply taken as $t=m\delta$ and $\delta\to 0$.
Then dNLS \eqref{eqn:dNLS}  and corresponding deformation equation \eqref{eqn:discrete_def_Frenet} 
and \eqref{eqn:discrete_coeffs_Frenet} yields the {\em semi-discrete NLS equation} or the {\em Ablowitz-Ladik equation} 
\cite{Ablowitz-Ladik:SiAM1976,Ablowitz-Ladik:SiAM1977}
\begin{equation}
\sqrt{-1}\frac{du_n}{dt} + \frac{u_{n+1} -2u_n + u_{n-1}}{\epsilon^2} + (u_{n+1} + u_{n-1})|u_n|^2=0 ,
\end{equation}
and the deformation equation of discrete space curves \cite{Doliwa-Santini:JMP,Hisakado-Wadati,Nakayama:JPSJ2007}
\begin{equation}
 \frac{d}{dt}\gamma_n = \frac{2}{\epsilon}\tan\frac{\kappa_n}{2} B_n.
\end{equation}
\end{enumerate}
\end{rem}
%%%%%%%%%%%%%%%%%%%%%%%%%%%%%%%%%%%%%%%%%%%%%%%%%%%%%%%%%%%%%%
%
%%%%%%%%%%%%%%%%%%%%%%%%%%%%%%%%%%%%%%%%%%%%%%%%%%%%%%%%%%%%%%
The dNLS flow \eqref{eqn:discrete_def_Frenet} and \eqref{eqn:discrete_coeffs_Frenet} implies the
deformation of Frenet frame as
\begin{equation}\label{eqn:discrete_deformation_Frenet}
\begin{split}
& \Phi_n^{m+1} = \Phi_n^m M^m_n,\\
&M_n^m = \frac{1}{\Gamma^m_{n + 1}}
\begin{bmatrix}
\left|\alpha^m_n\right|^2 - \left|\beta^m_n\right|^2 &
2 \Re \left(\alpha^m_n \beta^m_n{}^\ast\right) &
- 2 \Im \left(\alpha^m_n \beta^m_n{}^\ast{}\right)\\
- 2 \Re \left(\alpha^m_n \beta^m_n\right) &
\Re \left({\alpha^m_n}^2 - {\beta^m_n}^2\right) &
- \Im \left({\alpha^m_n}^2 + {\beta^m_n}^2\right)\\
- 2 \Im \left(\alpha^m_n \beta^m_n\right) &
\Im \left({\alpha^m_n}^2 - {\beta^m_n}^2\right) &
\Re \left({\alpha^m_n}^2 + {\beta^m_n}^2\right)
\end{bmatrix}\in\mathrm{SO}(3) ,
\end{split}
\end{equation}
where $\alpha_n^m,\beta_n^m\in\mathbb{C}$ are given by
\begin{equation}\label{eqn:alpha_beta}
\begin{split}
 \alpha_n^m
&=
\sqrt{- 1} \frac{\delta}{\epsilon^2}
\left[ \left(1-\sqrt{-1}\frac{\epsilon^2}{\delta}\right) 
     - \left(1 + \epsilon^2 u_{n + 1}^m u_n^{m+1}{}^*\right) \Gamma_{n + 1}^m\right] 
e^{\frac{\sqrt{-1}}{2}(\Lambda_n^{m+1} - \Lambda_n^m)},\\
\beta_n^m
&=
\sqrt{- 1} \frac{\delta}{\epsilon}
\left(u_n^{m+1} - u_{n + 1}^m\right) \Gamma_{n + 1}^m
e^{- \frac{\sqrt{- 1}}{2}(\Lambda_n^m + \Lambda_n^{m+1})},
\end{split}
\end{equation}
respectively. Here, $*$ means the complex conjugate. The Frenet-Serret formula
\eqref{eqn:fulldiscrete_Frenet-Serret} and the deformation equation
\eqref{eqn:discrete_deformation_Frenet} can be transformed to the $\mathrm{SU}(2)$ version by the
standard correspondence of $\mathrm{SO}(3)$ and $\mathrm{SU}(2)$ as
\begin{equation}
\begin{split}
& \phi_{n+1}^m = \phi_n^m L_n^m,\quad L_n^m=
 \begin{bmatrix}\medskip
\cos \frac{\kappa^m_{n + 1}}{2}
e^{- \frac{\sqrt{- 1} }{2}\nu^m_{n + 1}} &
- \sin \frac{\kappa^m_{n + 1}}{2}
e^{- \frac{\sqrt{- 1} }{2}\nu^m_{n + 1}}\\
\sin \frac{\kappa^m_{n + 1}}{2}
e^{\frac{\sqrt{- 1} }{2}\nu^m_{n + 1}} &
\cos \frac{\kappa^m_{n + 1}}{2}
e^{\frac{\sqrt{- 1} }{2}\nu^m_{n + 1}}
\end{bmatrix},\\
& \phi_{n}^{m+1} = \phi_n^m M_n^m,\quad M_n^m=
\frac{1}{\sqrt{\Gamma^m_{n + 1}}}
\begin{bmatrix}
\alpha^m_n & \beta^m_n\\
- \beta^m_n{}^\ast & \alpha^m_n{}^\ast
\end{bmatrix},
\end{split}
\end{equation}
which is known as the {\em Lax pair} of dNLS \cite{Ablowitz-Ladik:SiAM1976,Ablowitz-Ladik:SiAM1977}. In fact, one can
verify that the compatibility condition $L_n^{m}M_{n+1}^m=M_{n}^mL_n^{m+1}$ yields dNLS \eqref{eqn:dNLS}.
%%%%%%%%%%%%%%%%%%%%%%%%%%%%%%%%%%%%%%%%%%%%%%%%%%%%%%%%%%%%%%
%
%%%%%%%%%%%%%%%%%%%%%%%%%%%%%%%%%%%%%%%%%%%%%%%%%%%%%%%%%%%%%%
\paragraph{Outline of the proof of Theorem \ref{thm:discrete_deformation}}
The first statement may be verified directly in principle, by computing
$\gamma_{n+1}^{m+1}-\gamma_n^{m+1}$ and its length from \eqref{eqn:discrete_def_Frenet},
\eqref{eqn:discrete_coeffs_Frenet} and the discrete Frenet-Serret formula
\eqref{eqn:fulldiscrete_Frenet-Serret} under the assumption that $u_n^{m+1}$ is determined by dNLS
\eqref{eqn:dNLS}. However, this computation is hopelessly complicated to carry out. To make it
feasible, we change the Frenet frame to a different frame used in
\cite{Hasimoto:JFM1972,Lamb:JMP1977}, which we call the {\em complex parallel frame} in this
paper. Let $F_n^m=[T_n^m,U_n^m,U_n^m{}^\ast]\in\mathrm{U}(3)$ be the complex parallel frame defined by
\begin{equation}
 U_n^m = \frac{e^{\sqrt{- 1} \Lambda_n^m}}{\sqrt{2}} \left(N_n^m + \sqrt{- 1} B_n^m\right).
\end{equation}
Note that it is related to the discrete Frenet frame $\Phi_n^m$ as
\begin{equation}\label{eqn:discrete_Frenet_complex_parallel}
F_n^m = \Phi_n^m
\begin{bmatrix}
1 & 0 & 0\\
0 & 1 & 1\\
0 & \sqrt{- 1} & - \sqrt{- 1}
\end{bmatrix}
\begin{bmatrix}
1 & 0 & 0\\
0 & \frac{e^{\sqrt{- 1} \Lambda_n^m}}{\sqrt{2}} & 0\\
0 & 0 & \frac{e^{-\sqrt{- 1} \Lambda_n^m}}{\sqrt{2}}
\end{bmatrix}.
\end{equation}
Then the complex curvature $u_n^m$ naturally arises in this framework;
the discrete Frenet-Serret formula \eqref{eqn:fulldiscrete_Frenet-Serret} and the deformation of
the discrete curve are rewritten in terms of $u_n^m$ as
\begin{equation}
 F_{n + 1}^m = F_n^m X_n^m,\ 
X_n^m =
{\footnotesize \frac{1}{1 + \epsilon^2 \left|u_{n + 1}^m\right|^2}
\begin{bmatrix}
1 - \epsilon^2 \left|u_{n + 1}^m\right|^2 &- \sqrt{2} \epsilon u_{n + 1}^m &- \sqrt{2} \epsilon u_{n + 1}^m{}^\ast\\[1mm]
\sqrt{2} \epsilon u_{n + 1}^m{}^\ast & 1 & - \epsilon^2 (u_{n + 1}^m{}^\ast)^2\\[1mm]
\sqrt{2} \epsilon u_{n + 1}^m & - \epsilon^2 (u_{n + 1}^m)^2 & 1
\end{bmatrix}},
\end{equation}
and 
\begin{equation}
\gamma_n^{m+1}
= \gamma_n^m
+ \frac{2 \delta^2}{\epsilon^3}
F_n^m
\begin{bmatrix}
- 1 + \left(1 + \epsilon^2 \left|u_n^m\right|^2\right) \Gamma_n^m\\[2mm]
{\textstyle \frac{\epsilon}{\sqrt{2}}
\left\{\left(1-{\scriptstyle  \sqrt{-1}}\frac{\epsilon^2}{\delta} \right)u_n^m{}^\ast 
- u_{n - 1}^{m+1}{}^\ast \left(1 + \epsilon^2 \left|u_n^m\right|^2\right) \Gamma_n^m\right\}}\\[3mm]
{\textstyle \frac{\epsilon}{\sqrt{2}}
\left\{\left(1+{\scriptstyle  \sqrt{-1}}\frac{\epsilon^2}{\delta} \right) u_n^m
 - u_{n - 1}^{m+1} \left(1 + \epsilon^2 \left|u_n^m\right|^2\right) \Gamma_n^m\right\}}
\end{bmatrix},
\end{equation}
respectively. The following lemma plays a crucial role in the proof:
%%%%%%%%%%%%%%%%%%%%%%%%%%%%%%%%%%
%
%%%%%%%%%%%%%%%%%%%%%%%%%%%%%%%%%%
\begin{lemma}\label{lem:isoperimetricity}
Let $\gamma_n^m\in\mathbb{R}^3$ be the family of discrete space curves given in Theorem
\ref{thm:discrete_deformation}. Then it follows that
\begin{equation}\label{eqn:isoperimetricity}
 |\alpha_n^m|^2 + |\beta_n^m|^2 = \Gamma_{n+1}^m.
\end{equation}
%provided that $\Gamma_{\pm\infty} = 1$ or $1+\frac{\epsilon^4}{\delta^2}$.
\end{lemma}
%%%%%%%%%%%%%%%%%%%%%%%%%%%%%%%%%%
%
%%%%%%%%%%%%%%%%%%%%%%%%%%%%%%%%%%
By using \eqref{eqn:isoperimetricity}, we have after long but straightforward calculations
\begin{equation}\label{eqn:T_m+1}
T_n^{m+1}=\frac{\gamma_{n + 1}^{m+1} - \gamma_n^{m+1}}{\epsilon}
=
F_n^m\;
\frac{1}{\Gamma_{n + 1}^m}
\begin{bmatrix}
\left|\alpha_n^m\right|^2 - \left|\beta_n^m\right|^2\\[1mm]
- \sqrt{2}
\alpha_n^m{}^\ast \beta_n^m{}^\ast e^{-\sqrt{- 1} \Lambda_n^m}\\[1mm]
- \sqrt{2}
\alpha_n^m \beta_n^m e^{ \sqrt{- 1} \Lambda_n^m}
\end{bmatrix},
\end{equation}
from which we obtain
\begin{displaymath}
 \left|\frac{\gamma_{n + 1}^{m+1} - \gamma_n^{m+1}}{\epsilon}\right|^2
= \frac{\left(\left|\alpha_n^m\right|^2 - \left|\beta_n^m\right|^2\right)^2
+ 4 \left|\alpha_n^m\right|^2 \left|\beta_n^m\right|^2}%
{{\Gamma_{n + 1}^m}^2}
= \left(\frac{\left|\alpha_n^m\right|^2 + \left|\beta_n^m\right|^2}
{\Gamma_{n + 1}^m}\right)^2
= 1.
\end{displaymath}
This proves the first statement. The second statement is proved as follows. Starting from
$T_n^{m+1}$ \eqref{eqn:T_m+1}, we have $B_n^{m+1}$ and $N_n^{m+1}$ in terms of $F_n^m$ by using
\eqref{eqn:discrete_Frenet_frame}. Then we obtain an expression of
$\Phi_n^{m+1}=[T_n^{m+1},N_n^{m+1},B_n^{m+1}]$ in terms of $F_n^m$, which can be rewritten as
$F_n^{m+1}=F_n^m Y_n^m$ with a certain matrix $Y_n^m\in\mathrm{U}(3)$ by using
\eqref{eqn:discrete_Frenet_complex_parallel}. This can be also transformed to the deformation
equation of discrete Frenet frame of the form $\Phi_n^{m+1}=\Phi_n^m M_n^m$ with $M_n^m$ given in
\eqref{eqn:discrete_deformation_Frenet}. Finally one can check that $\Phi_n^{m+1}$ satisfies the
discrete Frenet-Serret formula \eqref{eqn:fulldiscrete_Frenet-Serret} for $\kappa_n^{m+1}$ and
$\nu_n^{m+1}$ determined from the complex curvature $u_n^{m+1}$. This completes the proof of Theorem
\ref{thm:discrete_deformation}.\qquad $\square$

%%%%%%%%%%%%%%%%%%%%%%%%%%%%%%%%%%
%
%%%%%%%%%%%%%%%%%%%%%%%%%%%%%%%%%%
\section{Explicit formulas} 
The formulation of NLS and dNLS flows in terms of the Frenet frame enables us to apply the theory
of integrable systems.  As an example, we here present explicit formulas of the NLS and dNLS flows
in terms of the $\tau$ functions. For the case of plane curves, see \cite{IKMO:KJM}. These
formulas are established based on the bilinear formalism in the theory of integrable systems by
applying suitable reductions and imposing complex structure to $\tau$ functions of the 2-component
KP hierarchy, but here we only show the results, leaving full derivations to the forthcoming publications.

For any $N\in\mathbb{N}$, we first introduce the following three determinants, a $2N\times 2N$ determinant
$\tau$, two $(2N+1)\times (2N+1)$ determinants $\sigma$ and $\rho$ as
\begin{equation}
\begin{scriptsize}
 \tau = \left|
\begin{array}{ccccccc}
 m_{11}^{(1)}&\cdots &m_{1N}^{(1)} &\ \ \vbl{4}\ \  &1 &   &\O \\[-1.5mm]
 \vdots      &\cdots &\vdots       &\ \ \vbl{4}\ \  &  &\ddots & \\
 m_{N1}^{(1)}&\cdots &m_{NN}^{(1)} &\ \ \vbl{4}\ \  &\O  & &1 \\
\multispan{7}\hblfil\\
-1           &       &\O           &\ \ \vbl{4}\ \  &m_{11}^{(2)} &\cdots &m_{1N}^{(2)}\\[-1.5mm]
             &\ddots &             &\ \ \vbl{4}\ \  &\vdots       &\cdots &\vdots\\
\O           &       & -1          &\ \ \vbl{4}\ \  &m_{N1}^{(2)} &\cdots &m_{NN}^{(2)}
\end{array}
\right|,
\end{scriptsize}
\end{equation}
\begin{equation}
\begin{scriptsize}
  \sigma = \left|
\begin{array}{ccccccccc}
 m_{11}^{(1)}&\cdots &m_{1N}^{(1)} &\ \ \vbl{4}\ \   &1 &       &\O & \ \ \vbl{4}\ \   & \varphi_1^{(1)}\\[-1.5mm]
 \vdots      &\cdots &\vdots       &\ \ \vbl{4}\ \   &  &\ddots &   & \ \ \vbl{4}\ \   & \vdots\\
 m_{N1}^{(1)}&\cdots &m_{NN}^{(1)} &\ \ \vbl{4}\ \   &\O&       &1  & \ \ \vbl{4}\ \   & \varphi_N^{(1)}\\
\multispan{9}\hblfil\\
-1           &       &\O           &\ \ \vbl{4}\ \   &m_{11}^{(2)} &\cdots &m_{1N}^{(2)} &\ \ \vbl{4}\ \   & 0\\[-1.5mm]
             &\ddots &             &\ \ \vbl{4}\ \   &\vdots       &\cdots &\vdots       &\ \ \vbl{4}\ \   & \vdots\\
\O           &       & -1          &\ \ \vbl{4}\ \   &m_{N1}^{(2)} &\cdots &m_{NN}^{(2)} &\ \ \vbl{4}\ \   & 0\\
\multispan{9}\hblfil\\
0            &\cdots & 0           &\ \ \vbl{4}\ \   & \varphi_1^{(2)} &\cdots & \varphi_N^{(2)}&\ \ \vbl{4}\ \   & 0
\end{array}
\right|,
\end{scriptsize}
\end{equation}
\begin{equation}
\begin{scriptsize}
  \rho= \left|
\begin{array}{ccccccccc}
 m_{11}^{(1)}&\cdots &m_{1N}^{(1)} &\ \ \vbl{4}\ \   &1 &       &\O & \ \ \vbl{4}\ \   & 0\\[-1.5mm]
 \vdots      &\cdots &\vdots       &\ \ \vbl{4}\ \   &  &\ddots &   & \ \ \vbl{4}\ \   & \vdots\\
 m_{N1}^{(1)}&\cdots &m_{NN}^{(1)} &\ \ \vbl{4}\ \   &\O&       &1  & \ \ \vbl{4}\ \   & 0\\
\multispan{9}\hblfil\\
-1           &       &\O           &\ \ \vbl{4}\ \   &m_{11}^{(2)} &\cdots &m_{1N}^{(2)} &\ \ \vbl{4}\ \   & \psi_1^{(1)}\\[-1.5mm]
             &\ddots &             &\ \ \vbl{4}\ \   &\vdots       &\cdots &\vdots       &\ \ \vbl{4}\ \   & \vdots\\
\O           &       & -1          &\ \ \vbl{4}\ \   &m_{N1}^{(2)} &\cdots &m_{NN}^{(2)} &\ \ \vbl{4}\ \   & \psi_N^{(1)}\\
\multispan{9}\hblfil\\
\psi_1^{(2)} &\cdots & \psi_N^{(2)}&\ \ \vbl{4}\ \   & 0&\cdots & 0&\ \ \vbl{4}\ \   & 0
\end{array}
\right|,
\end{scriptsize}
\end{equation}
where $\O$ is the empty block. Then the formulas for NLS flow on smooth curves and dNLS flow on
discrete curves are obtained by choosing the entries of determinant as follows:
\paragraph{NLS flow on smooth curves}
We choose the entries of determinants as
\begin{equation}
\begin{split}
& \psi_i^{(1)} = 1,\quad \psi_i^{(2)} = -\left(-\frac{1}{p_i^*}\right)^n e^{\xi_i^*},\\
& \varphi_i^{(1)} = p_i^n e^{\xi_i},\quad \varphi_i^{(2)} = -1,\\
& m_{ij}^{(1)} = -\frac{\varphi_i^{(1)}\psi_j^{(2)}}{p_i+p_j^*},\quad 
m_{ij}^{(2)} = \frac{1}{p_i^*+p_j},\\
&\xi_i = p_i x - \sqrt{-1}p_i^2t+\frac{1}{p_i}z + \xi_i^{(0)},\quad p_i,\ \xi_i^{(0)}\in\mathbb{C},
\end{split}
\end{equation}
so that we write $\tau=\tau_n(x,t;z)$, $\sigma=\sigma_n(x,t;z)$, $\rho=\rho_n(x,t;z)$. Here,
$n$ and $z$ are regarded as auxiliary variables. Putting
\begin{equation}
 F = \tau_0,\quad G= -\rho_0,\quad h=-\sigma_{-2},%\quad h^* = \rho_2,
\end{equation}
we have:
\begin{thm}[Explicit formula for NLS flow]\footnote{The $N$-soliton solution for the tangent vector has been constructed by using the bilinear formalism in \cite{Fukumoto_Miyazaki:JPSJ1986}.}\label{thm:explict_formula_smooth}\hfill
\begin{enumerate}
 \item Let $u=u(x,t)\in\mathbb{C}$ be 
\begin{equation}
 u = 2\frac{G}{F}.
\end{equation}
Then $u$ satisfies NLS \eqref{eqn:NLS}.
 \item Let $\gamma=\gamma(x,t)\in\mathbb{R}^3$ be
\begin{equation}
 \gamma = \left[
\begin{array}{c}
{\displaystyle \frac{h + h^*}{F}}\\[4mm]
{\displaystyle \frac{1}{\sqrt{-1}}\frac{h - h^*}{F}}\\[4mm]
{\displaystyle 2\frac{\partial}{\partial z}(\log F) - x}
\end{array}
\right].
\end{equation}
Then $\gamma$ satisfies the Frenet-Serret formula \eqref{eqn:Frenet-Serret} and the deformation equation 
\eqref{eqn:continuous_binormal_flow}.
\end{enumerate}
\end{thm}
\begin{figure}[ht]
 \begin{center}
\includegraphics[bb=0 0 181 432,scale=0.25]{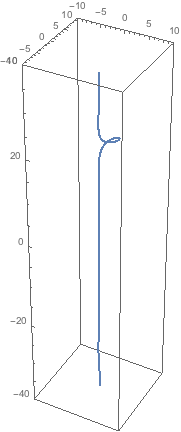}\hskip40pt
  \includegraphics[bb=0 0 181 432,scale=0.25]{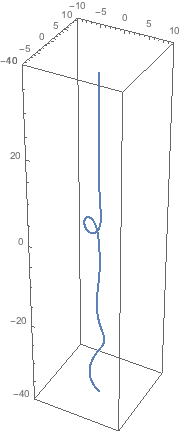}\hskip40pt
  \includegraphics[bb=0 0 181 432,scale=0.25]{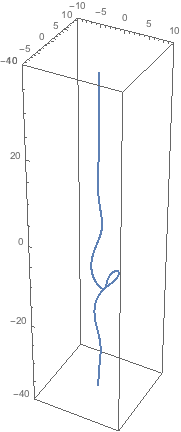}\hskip40pt
  \includegraphics[bb=0 0 181 432,scale=0.25]{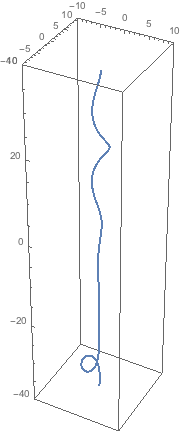}
 \end{center}
\caption{Interaction of loops of smooth curve by NLS flow obtained from Theorem \ref{thm:explict_formula_smooth}.}
\end{figure}
%%%%%%%%%%%%%%%%%%%%%%%%%%%%%%%%%%%%%%%%%%%%%%%%%%%%%
%
%%%%%%%%%%%%%%%%%%%%%%%%%%%%%%%%%%%%%%%%%%%%%%%%%%%%%
\paragraph{dNLS flow on discrete curves}
We choose the entries of determinants as
\begin{equation}
\begin{split}
& {\textstyle \varphi_i^{(1)} = p_i^{-n} e^{\zeta_i} (1-ap_i)^{-m}(1-cp_i)^{-r},\ 
\varphi_i^{(2)} = -\left(1-\frac{a}{p_i}\right)^{m}\left(1-\frac{1}{cp_i}\right)^{s}},\\
&{\textstyle  \psi_i^{(1)} = (1-ap_i^*)^{-m}\left(1-\frac{p_i^*}{c}\right)^{-s},\ 
\psi_i^{(2)} = -(p_i^*)^{-n} e^{\zeta_i^*} \left(1-\frac{a}{p_i^*}\right)^{m}\left(1-\frac{c}{p_i^*}\right)^{r},}\\
& 
m_{ij}^{(1)}= - \frac{\varphi_i^{(1)}\psi_j^{(2)}}{p_i-\frac{1}{p_j^*}},\quad
m_{ij}^{(2)}= - \frac{\psi_i^{(1)}\varphi_j^{(2)}}{p_i^*-\frac{1}{p_j}},\quad
e^{\zeta_i} = e^{\frac{1}{2}\frac{1+cp_i}{1-cp_i}z},\\
&\hskip60pt  a\in\mathbb{R},\quad p_i,\ c\in\mathbb{C},\quad |c|=1,
\end{split}
\end{equation}
so that we write $\tau = \tau_n^m(r,s;z)$, $\sigma=\sigma_n^m(r,s;z)$, $\rho=\rho_n^m(r,s;z)$ with $r$, $s$
and $z$ being auxiliary variables. Putting
\begin{equation}
F_n^m = \tau_n^m(0,0;z),\quad G_n^m = \rho_n^m(0,0;z),\quad h_n^m = c^{-n}\sigma_n^m(1,-1;z),
\end{equation}
\begin{equation}
a= \left(1+\frac{\epsilon^4}{\delta^2}\right)^{\frac{1}{2}},\quad 
c=\frac{1-\sqrt{-1}\frac{\epsilon^2}{\delta}}{a},
\end{equation}
we have:
%%%%%%%%%%%%%%%%%%%%%%%%%%%%%%%%%%%%%%%%%%%%%%%%%%%%%
%
%%%%%%%%%%%%%%%%%%%%%%%%%%%%%%%%%%%%%%%%%%%%%%%%%%%%%
\begin{thm}[Explicit formula for dNLS flow]\label{thm:explict_formula_fulldiscrete}\hfill
\begin{enumerate}
 \item Let $u_n^m\in\mathbb{C}$ be
\begin{equation}
 u_n^m = \frac{(-1)^m c^{-n-2m}}{\epsilon}\frac{G_n^m}{F_n^m},\quad 
\Gamma_n^m = \frac{2a}{c+\frac{1}{c}}\frac{F_{n-1}^{m+1}F_{n}^{m}}{F_{n}^{m+1}F_{n-1}^{m}}.
\end{equation}
Then $u_n^m$ satisfies dNLS \eqref{eqn:dNLS}.\\
\item Let $\gamma_n^m\in\mathbb{R}^3$ be
\begin{equation}
 \gamma_n^m = \epsilon \left[
\begin{array}{c}
{\displaystyle (-1)^m\frac{h_n^m + h_n^m{}^*}{F_n^m}}\\[2mm]
{\displaystyle \frac{(-1)^m}{\sqrt{-1}}\frac{h_n^m - h_n^m{}^*}{F_n^m}}\\[2mm]
{\displaystyle 2\frac{\partial}{\partial z}(\log F_n^m) - n - 2m}
\end{array}
\right].
\end{equation}
Then $\gamma_n^m$ satisfies the Frenet-Serret formula \eqref{eqn:fulldiscrete_Frenet-Serret} and the deformation equation 
\eqref{eqn:discrete_def_Frenet}, \eqref{eqn:discrete_coeffs_Frenet}.
\end{enumerate}
\end{thm}
%%%%%%%%%%%%%%%%%%%%%%%%%%%%%%%%%%%%%%%%%%%%%%%%%%%%%
%
%%%%%%%%%%%%%%%%%%%%%%%%%%%%%%%%%%%%%%%%%%%%%%%%%%%%%
\begin{figure}[ht]
 \begin{center}
\includegraphics[bb=100 150 300 700,scale=0.18,clip]{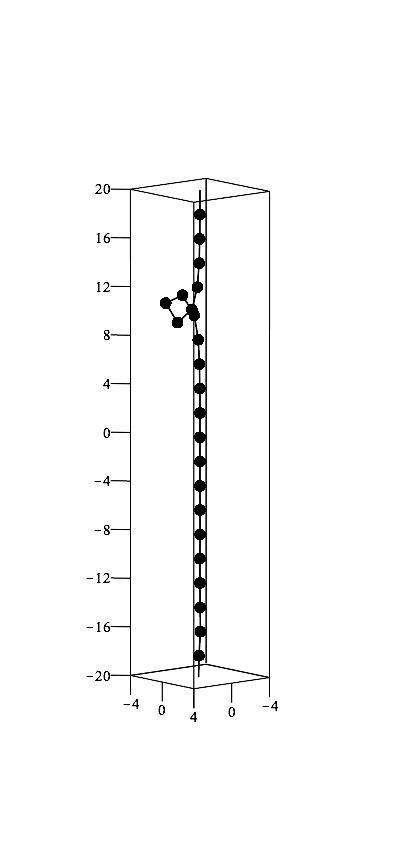}\hskip40pt
\includegraphics[bb=100 150 300 700,clip,scale=0.18]{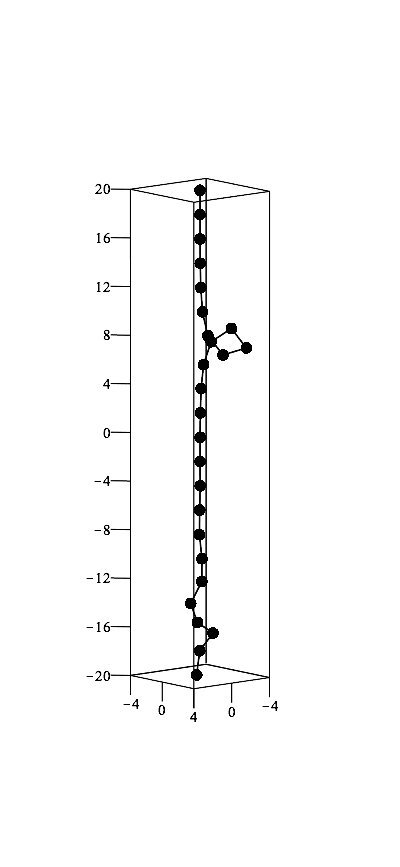}\hskip40pt
\includegraphics[bb=100 150 300 700,clip,scale=0.18]{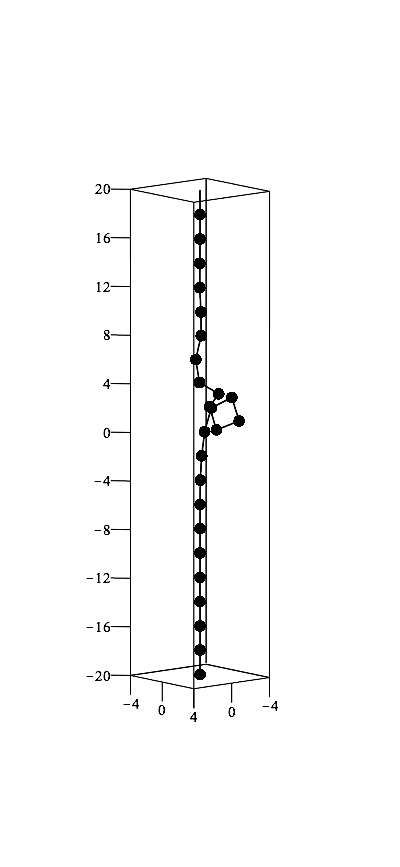}\hskip40pt
\includegraphics[bb=100 150 300 700,clip,scale=0.18]{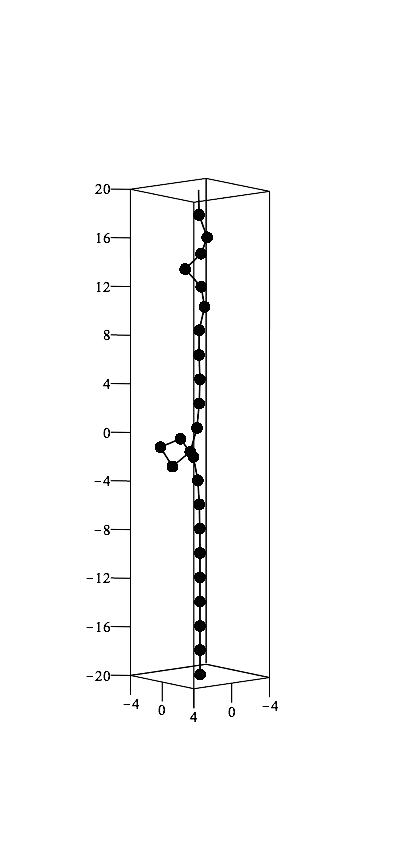}
 \end{center}
\caption{Interaction of loops of discrete curve by dNLS flow obtained from Theorem \ref{thm:explict_formula_fulldiscrete}.}
\end{figure}

\end{document}